%Paper: hep-th/9303040
%From: vadim@bolvan.ph.utexas.edu (Vadim S. Kaplunovsky)
%Date: Mon, 8 Mar 93 04:18:57 -0600

\input phyzzx

%&phyzzx
%macropackage=phyzzx
%\input phyzzx
%%%%%%%%%%%%%%%%%%%%%%%%%%%%%%%%%%%%%%%%%%%%%%%%
% Format switch
%
% uncomment ONE of the following definitions
%
 \let\SELECTOR=P      % un-reduced Preprint format
%\let\SELECTOR=R      % Reduced preprint format
%\let\SELECTOR=W      % World Scientific format
%
% the Reduced format may not work with some DVI drivers
%%%%%%%%%%%%%%%%%%%%%%%%%%%%%%%%%%%%%%%%%%%%%%%%
%
% Formatting macros
%
\expandafter\ifx\csname SELECTOR\endcsname\relax \let\SELECTOR=P \fi
% P is the default format
\if W\SELECTOR
    \nopagenumbers
    \vsize=51pc
    \hsize=36pc
    \baselineskip=14pt
    \parskip = 4pt plus 2pt minus 1pt
    \twelvepoint
    \headskip = 4mm plus 0.5mm
    \frontpageskip = 6mm  plus 1mm
    \referenceminspace=8pc
    \def\abstract{\par\vskip\frontpageskip \smallskip
        \centerline{ABSTRACT} \vskip\headskip
        \baselineskip 12pt
        \leftskip 3pc \rightskip 3pc
        \noindent }
    \def\titlepage{\par\begingroup\tenpoint }
    \def\endtitlepage{\par \endgroup \vskip\frontpageskip \smallskip }
    \def\titleupset#1{{\tenbf\uppercase{#1}}}
    \def\talk{talk}
\else
    \hsize=175mm
    \vsize=225mm
    \hoffset=-5mm
    \voffset=0mm
    \def\titleupset{\relax}
    \def\endtitlepage{\par
        \insert\footins{\floatingpenalty=20000 \vbox{\hbox{}\medskip
            \ialign{##\hfil\cr \the\Pubnum\crcr\the\date\crcr }
            \vskip -1.5\baselineskip }}
        \vfil \eject }
    
    \def\talk{article}
\fi

\newdimen\doublewidth
\newinsert\LeftPage
\count\LeftPage=0
\dimen\LeftPage=\maxdimen
\if R\SELECTOR
    \mag=833
    \vsize=161.5truemm
    \voffset=0truemm
    \hsize=117truemm
    \hoffset=-7truemm
    \doublewidth=250truemm
    \def\PageBox{\vbox{\makeheadline \pagebody \makefootline }}
    \output={\ifvoid\LeftPage \insert\LeftPage{\floatingpenalty 20000 \PageBox}
        \else \shipout\hbox to\doublewidth{%
            \box\LeftPage \hfil \PageBox }\fi
        \advancepageno
        \ifnum\outputpenalty>-20000 \else \dosupereject \fi }
    \message{Warning: some DVI drivers cannot handle reduced output!!!}
\fi

%
%%%%%%%%%%%%%%%%%%%%%%%%%%%%%%%%%%%%%%%%%%%%%%%%%
% general settings and macros
%
%\overfullrule=0pt
%\draft
\interdisplaylinepenalty=10000
\def\del{\partial}
 
\def\Re{\mathop{\rm Re}\nolimits}

\def\refmark{\attach}
\def\vev#1{\mathopen\langle #1\mathclose\rangle }
%\def\eg{\hbox{ e.g.}}
%\def\ie{\hbox{ i.e.}}
%
%%%%%%%%%%%%%%%%%%%%%%%%%%%%%%%%%%%%%%%%%%%%%%%
% indices, fields, couplings, etc.
%
\def\ib{{\bar\imath}}
\def\jb{{\bar\jmath}}
\def\Ib{{\bar I}}
\def\Jb{{\bar J}}
\def\ind{a}
\def\Fb{\overline{F}}
\def\Mb{\overline M}
\def\matter{Q}
\def\matterb{\smash{\overline\matter}\vphantom{\matter}}
\def\hmat{A}
\def\modul{\Phi}
\def\modulb{\smash{\overline\Phi}\vphantom{\Phi}}
\def\Kmod{{\hat{K}}}
\def\Zh{H}
\def\Wtree{W^{\rm (tree)}}

\def\Weff{W^{\rm (eff)}}
\def\Wh{W^{\rm (ind)}}
\def\Wmod{{\hat W}}
\def\Veff{V^{\rm (eff)}}

\def\Gobs{G^{\rm (obs)}}
\def\Ghid{G^{\rm (hid)}}
\def\Dmod{D}
\def\mpl{M_{\rm Pl}}
\def\mgrav{m_{3/2}}
\def\mweak{M_{\rm W}}
\def\ms{M_{\rm hid}}
\def\mb{B}
\def\muh{\mu}
\def\muind{\tilde{\mu}}
\def\Y{Y}
\def\Ys{\tilde{Y}}
\def\cano{{\bf c}}
\def\cren{{\bf b}}
%
%%%%%%%%%%%%%%%%%%%%%%%%%%%%%%%%%%%%%%%%%%%%%%%%
% macros for references
%
\def\us#1{\underline{#1}}
\def\ldf{\REF}
\def\nup#1({Nucl.\ Phys.\ $\us {B#1}$\ (}
\def\plt#1({Phys.\ Lett.\ $\us  {#1}$\ (}
\def\cmp#1({Comm.\ Math.\ Phys.\ $\us  {#1}$\ (}
\def\prp#1({Phys.\ Rep.\ $\us  {#1}$\ (}
\def\prl#1({Phys.\ Rev.\ Lett.\ $\us  {#1}$\ (}
\def\prv#1({Phys.\ Rev.\ $\us  {#1}$\ (}
\def\mplt#1({Mod.\ Phys.\ \Let.\ $\us  {#1}$\ (}
\def\tit#1|{{\it #1},\ }
\def\K{K\"ahler}
%
%
%%%%%%%%%%%%%%%%%%%%%%%%%%%%%%%%%%%%%%%%%%%%%
% references
%
\ldf\MSSM{For a review, see for example,
H.-P.~Nilles, \prp C110 (1984) 1 and references therein.}
\ldf\gauginoc{For a review, see for example,
 D.~Amati, K.~Konishi, Y.~Meurice, G.~Rossi and
G.~Veneziano, \prp  162 (1988) 169;
 H.-P.~Nilles,
Int. J. Mod. Phys. $\us{A5}$ (1990) 4199;
J. Louis, in  Proceedings of the 1991 DPF Meeting, Vancouver, B.C., Canada,
(World Scientific,
Singapore, 1992), and references therein.}
\ldf\BF{
R.~Barbieri, S.~Ferrara and C.~Savoy, \plt 119B (1982) 343.}
\ldf\SW{S.K.~Soni and H.A.~Weldon, \plt 126B (1983) 215.}
\ldf\GM{G.F.~Giudice and A.~Masiero, \plt 206B (1988) 480.}
\ldf\KN{J.E.~Kim and H.-P.~Nilles, \plt 138B (1984) 150;
E.J.~Chun,  J.E.~Kim,  H.-P.~Nilles, \nup370 (1992) 105. }
\ldf\dindrsw{J.P.~Derendinger, L.E.~Ib\'a\~nez and
  H.P.~Nilles, \plt  155B (1985) 65;
M.~Dine, R.~Rohm, N.~Seiberg and
  E.~Witten, \plt  156B (1985) 55.}
\ldf\krasnikov{N.V.~Krasnikov, \plt  193B (1987) 37;
L.~Dixon, in  Proceedings of  the A.P.S. DPF Meeting, Houston, 1990;
J.A.~Casas, Z.~Lalak, C.~Mu\~noz and G.G.~Ross, \nup347 (1990) 243;
T.~Taylor, \plt B252 (1990) 59.}
\ldf\FILQ{
A.~Font, L.E.~Ib\'a\~nez, D.~L\"ust and F.~Quevedo,
\plt  B245 (1990) 401.}
\ldf\Mbreak{
S.~Ferrara, N.~Magnoli, T.~Taylor and G.~Veneziano, \plt
 B245 (1990) 409;
H.-P.~Nilles and M.~Olechowski, \plt B248 (1990)
268;
P.~Bin\'etruy and M.K.~Gaillard, \nup358 (1991) 121.}
\ldf\CFILQ{M.~Cveti\v c, A.~Font, L.E.~Ib\'a\~nez, D.~L\"ust and
F.~Quevedo, \nup361 (1991) 194.}
\ldf\IL{L.~Ib\'a\~nez and D.~L\"ust, \nup382 (1992) 305.}
\ldf\CCMa{B.~de Carlos, J.A.~Casas and C.~Mu\~noz, CERN preprint
CERN-TH.6436/92.}
\ldf\CCMb{B.~de Carlos, J.A.~Casas and C.~Mu\~noz, CERN preprint
CERN-TH.6681/92.}
\ldf\KL{V.~Kaplunovsky and J.~Louis, to appear.}
\ldf\noscale{E.~Cremmer, S.~Ferrara, C.~Kounnas and D.V.~Nanopoulos,
    \plt 133B (1983) 61.}
\ldf\danomaly{
M.A.~Shifman and A.I.~Vainshtein,
   \nup277 (1986) 456 and $\us{B359}$ (1991) 571;
J.~Louis, in  Proceedings of the Second International Symposium on Particles,
Strings and Cosmology, eds.~P.~Nath and S.~Reucroft, (World Scientific,
Singapore, 1992);
G.~Cardoso Lopes and B.~Ovrut, \nup369 (1992) 351;
J.P.~Derendinger, S.~Ferrara, C.~Kounnas and F.~Zwirner, \nup372 (1992) 145.
}
\ldf\GG{L.~Girardello and M.T.~Grisaru, \nup194 (1982) 65.}
\ldf\EGHRZ{See for example J.~Ellis, J.~F.~Gunion, H.~Haber, L.~Roszkowski
and F. Zwirner, \prv D39 (1989) 844, and references therein.}
\ldf\DKLa{L.J.\ Dixon, V.S.\ Kaplunovsky and J.\ Louis,
  \nup329 (1990) 27.}
\ldf\AGNT{I.~Antoniadis, E.~Gava, K.~Narain and T.~Taylor,
 Northeastern preprint NUB-3057, Sep.~1992.}
\ldf\DKLANT{H.-P.~Nilles, \plt  180B (1986) 240;
L.~Dixon, V.~Kaplunovsky and J.~Louis, \nup355 (1991) 649;
I.~Antoniadis, K.~Narain and T.~Taylor, \plt B267 (1991) 37.}
\ldf\BLM{R.~Barbieri, J.~Louis and M.~Moretti, CERN preprint in preparation.}
\ldf\BG{R.~Barbieri and G.F.~Giudice, \nup306 (1988) 63.}
%
%%%%%%%%%%%%%%%%%%%%%%%%%%%%%%%%%%%%%%%%%%%%%%%%%%%%%%%%%%%%%%%%%%%%%%%%%
% Paper begins
%%%%%%%%%%%%%%%%%%%%%%%%%%%%%%%%%%%%%%%%%%%%%%%%%%%%%%%%%%%%%%%%%%%%%%%%%
\Pubnum={CERN--TH.6809/93\crcr UTTG--05--93}
\date={}
\titlepage
\date={February 1993}
\title{\titleupset{Model-Independent Analysis of Soft Terms\break
    in Effective Supergravity and in String Theory}%
  \if W\SELECTOR
  \foot{Talk presented by Jan Louis.}%
  \fi }
\author{\titleupset{Vadim~S. Kaplunovsky}%
  \if W\SELECTOR
  \else
  \foot{Research supported in part by the NSF under grant \#PHY--9009850
     and by the Robert A. Welch Foundation.}%
  \fi }
\address{Theory Group, Physics Department,
    University of Texas, Austin, TX~78712, USA}
\andauthor{\titleupset{Jan Louis}}
\address{CERN, Theory Division, CH-1211, Geneva 23, Switzerland}

\abstract
We discuss supersymmetry breakdown in effective supergravities
such as emerge in
%which emerge as
the low-energy limit of superstring theory.
Without specifying the precise trigger of the breakdown,
we analyse the soft parameters in the Lagrangian of the supersymmetrized
Standard Model.
\endtitlepage
%%%%%%%%%%%%%%%%%%%%%%%%%%%%%%%%%%%%%%%%%%%%%%%%%%%%%%%%%%%%%%%%%%%%%%%%

The Standard Model (SM) of particle interactions enjoys overwhelming
phenomenological success,
but it does not account for the gravitational
interactions  nor does it
explain the origin or naturalness of the electroweak scale $\mweak\ll\mpl$.
These theoretical problems result in a  belief
that the high-energy physics should be described
by a supergravity --- a locally supersymmetric quantum field theory that
contains  gravity, the SM, and perhaps some `hidden'
interactions in which known particles do not  participate.
The hierarchy of mass scales can be
naturally explained if the supersymmetry is exact at high energies
but becomes {\it spontaneously} broken, above $\mweak$,  by a non-perturbative
mechanism.
At low energies, this mechanism should decouple from the observable physics
and the supersymmetry would appear to be broken by explicit soft terms in
the effective low-energy Lagrangian.
{}From the low-energy point of view, these soft terms --- which include
the masses of the super-partners of all  known particles ---
are simply independent input parameters, just like the gauge and Yukawa
couplings of the SM, but from the high-energy point of view,
they are calculable in terms of  supergravity couplings.
\refmark{\MSSM}

Because of its non-renormalizability,
supergravity itself has to be thought of as an effective theory,
valid below the Planck scale $\mpl$.
Currently, the best candidate for a consistent theory governing the physics of
energies
$O(\mpl)$ and beyond is the heterotic string;
unfortunately, our present understanding of this theory is rather limited.
The state-of-the-art string theories are essentially series of world-sheet
topologies analogous to series of Feynman diagrams for quantum
field theories.
At this perturbative level,
the heterotic string has a large class of vacua that
lead to effective $(d=4,N=1)$ locally-supersymmetric field theories
at  energies below $\mpl$
and, at least in principle,
we know how to derive the Lagrangians of those effective theories.
Alas, the non-perturbative properties of string
theory are largely out of reach;
in particular, we have no `stringy' mechanisms
for the non-perturbative spontaneous breakdown of supersymmetry
or for the selection of the true string vacuum from the multitude of
perturbative
candidates.
Instead, one generally assumes that the dominant non-perturbative
effects originate at energies well below $\mpl$ and are thus
describable in terms of an effective field theory.
Gaugino condensation in an asymptotically free hidden sector of the effective
supergravity is a prime example of this mechanism.\refmark{\gauginoc}

In this \talk\ we pursue a different approach:
We do not  specify the precise trigger of supersymmetry breaking,
but instead parametrize its effects under several mild assumptions.
This allows us to make use of the `perturbative knowledge'
of the effective supergravity without relying upon our limited
knowledge of the non-perturbative physics.
In the same spirit, we  summarize the properties of the soft
% supersymmetry-breaking
terms generated under those assumptions.
Surprisingly, some of their features  are quite general and do
not depend upon a specific string vacuum; yet they
lead to distinct signals at $\mweak$.
In fact, some properties
apply to all effective supergravity theories,
regardless of the nature of the ultimate unified theory.
Many of our observations first appeared in the context of
gaugino condensation or within a specific class of string vacua
and our presentation relies heavily upon references \MSSM--\KL.
\if W\SELECTOR
    (See also contribution of D.~L\"ust in these proceedings.)
\fi
The main point of this \talk\ is to stress the generality of the
statements that can be made about the soft parameters.

Our starting point is
a phenomenologically acceptable effective $N=1$ supergravity theory
that also enjoys some features suggested
by generic properties of the $(d=4,N=1)$ superstring vacua.
In particular, we focus on
supergravities that have both
an `observable' sector --- a supersymmetric extension of the SM ---
and a `hidden' sector.
We do not insist upon `minimality' of the observable sector:
The observable gauge group $\Gobs$ may be extended beyond the
Standard $SU(3)\times SU(2)\times U(1)$, \eg\  by
adding an extra $U(1)$ factor.
Similarly, the chiral superfields $\matter^I$ of the observable sector
must include all the quark, lepton and Higgs superfields of the Minimal
Supersymmetric Standard Model (MSSM),\refmark{\MSSM}
but they may also include
additional particles with $O\rm(1\,TeV)$ masses.
The hidden sector generally has a large gauge group $\Ghid$ of
its own, as well as some chiral superfields $\hmat^\alpha$ that transform
non-trivially under $\Ghid$.
In this \talk, we leave the hidden sector completely generic and even allow
for some of the $\hmat^\alpha$ to be charged with respect to both $\Ghid$
and $\Gobs$.
However, in order to have strong non-perturbative effects in the hidden
sector, some of the factors of $\Ghid$ ought to be asymptotically free.

Furthermore, we expect to have a set of chiral moduli superfields $\modul^i$,
which are exact flat directions of the scalar potential
%and remain exactly flat
to all orders of perturbation theory.
\foot{In string theory, vacuum expectation values (VEVs) $\vev{\modul^i}$
    parametrize continuous families of the string vacua, hence the name
    `moduli'.
    From the effective supergravity point of view,
    the dilaton  $S$ ---  whose VEV is also
    undetermined to all orders of string perturbation theory ---
    is similar to the moduli, and for the time being,
    we treat $S$ as one of the $\modul^i$.
    We  return to the special properties of the dilaton later
    in this \talk.}
The non-perturbative effects in the hidden sector
generally lift this exact flatness.
They can also lead to a spontaneous breakdown of  supersymmetry;
this is signalled by a non-zero value of an $F$-term or a $D$-term,
which are the VEVs of
the auxiliary components of chiral and vector superfields.
In order to proceed we have to make a few assumptions
about the nature of the non-perturbative potential,
 the structure of the hidden sector and the  induced $F/D$-terms:

\pointbegin
The non-perturbative  potential has a stable
minimum with  no `runaway' directions.
(If this does not hold, the model should be re-analysed
by expanding around a different vacuum.)

\point
All hidden fields --- both gauge and matter --- become massive or confined
at an intermediate energy scale
$\ms$ well below $\mpl$ but well above $\mweak$ ($\mweak\ll\ms\ll\mpl$).
Some of the $\hmat^\alpha$ may acquire a VEV but
$\Gobs$ remains unbroken.

\point
By itself, the hidden sector does not break supersymmetry,
\ie\  $\vev{F^\alpha}=\vev{D^a}=0$ for $a\in\Ghid$.
(Relaxing this assumption goes beyond the scope of this
\talk\ but is certainly worth while  investigating.)

\noindent
These three assumptions allow us to integrate out the entire hidden sector
while  maintaining manifest supersymmetry of the remaining effective theory.
This induces an effective potential $\Veff(\modul)$ for
the moduli scalars, which we  assume to have the following features:

\point
$\Veff(\modul)$ has a stable minimum, without flat directions.

\point
At that minimum, $\Veff(\vev\modul)=0$, \ie\  the effective cosmological
constant vanishes (in some crude approximation).

\point
Some of the $\vev{F^i}$  in the moduli direction do not vanish,
 \ie\ supersymmetry is spontaneously broken in the moduli sector.

\point
The effect of this spontaneous
%supersymmetry
breakdown on the hidden sector
itself does not lead to additional {\sl large} VEVs of the hidden scalars.
%Similar to the assumption 3,
(It might also  be worth while to relax this assumption
and investigate the effects of the resulting feedback.)

\noindent
Some of our assumptions, in particular 1--3, are based upon our knowledge
of gaugino condensation, but we expect them to
be of more general validity.
On the other hand, the assumptions 4--7 are not so easily satisfied.
Stabilizing the dilaton's VEV is a generic problem of string theory and one
known solution  involves two or more independent
condensates.\refmark{\krasnikov,\CCMa}
The problem of a  vanishing cosmological constant is even more severe;
various possibilities have been investigated in refs.~\noscale\ and \CFILQ.
We refrain from further discussing  the validity of our assumptions;
instead  we analyse their effects  upon the  observable sector of the theory.

The Lagrangian of an effective  supergravity is completely specified in terms
of the (moduli-dependent) gauge couplings, the \K\ function $K$ and the
superpotential $W$.
The latter is a holomorphic function of the chiral superfields and does not
suffer from renormalization in any order of  perturbation theory,
but integrating out the hidden sector  introduces
non-perturbative corrections.
Thus, the superpotential of the effective theory for the moduli $\modul^i$
and the observable chiral superfields $\matter^I$
generally looks like  $W=\Wtree + \Wh$, where
$$
\Wtree(\modul,\matter)\ =\
\coeff13 \Ys_{IJL}(\modul)\, \matter^I \matter^J \matter^K\ +\ \cdots
\eqn\Wexpansion
$$
is the classical superpotential and
$$
\Wh(\modul,\matter)\ =\ \Wmod(\modul)\
+\ \coeff12 \muind_{IJ}(\modul)\, \matter^I \matter^J\ +\ \cdots
\eqn\Whexpansion
$$
summarizes the effects of integrating out
the hidden sector.
The $\cdots$ in both formulas stand for terms of higher order in $Q$
whose coefficients are suppressed by negative powers of $\mpl$.
Generally,  supersymmetric mass terms are absent from the
classical superpotential,
\foot{In perturbative string theory, most states  are either exactly massless
    or superheavy, and in the latter case they do not belong in the effective
    theory.
    However, it is possible for certain string states to have
    $\modul$-dependent masses that can vary all the way between
    $O(\mpl)$ and $0$, and in particular could be light in some region
    of the moduli space.
    (We would like to thank M.~Cveti\v c for a discussion of this point.)
    In this case, the combined $W(\modul,\matter)$ would have two kinds
    of mass terms, of very different origins, but phenomenologically
    indistinguishable from each other.}
but are induced by the hidden sector.
Yukawa-like cross-couplings
$O(\mpl^{1-n})\hmat^n\matter^2 \subset W$ (all indices suppressed)
between the observable and the hidden chiral superfields generate
$\muind_{IJ}$ if the hidden scalars acquire
% a VEV
VEVs
due to  non-perturbative effects.
Generically, $\vev{\hmat^\alpha}\sim\ms$ --- the characteristic
scale of the non-perturbative dynamics of the hidden sector,
--- which leads to $\muind_{IJ}=O(\ms^n/\mpl^{n-1})$.\refmark{\KN}

The \K\ function $K$  is a gauge-invariant real analytic
function of the chiral superfields  and is responsible for
their kinetic terms and $\sigma$-model interactions.
Expanding  in powers of $\matter^I$ and $\matterb^\Ib$, we have
$$
K\ =\ \kappa^{-2} \Kmod(\modul,\modulb)\
+\ Z_{\Ib J}(\modul,\modulb)\, \matterb^\Ib   \matter^J\ +\ \left(\coeff12
\Zh_{I J}(\modul,\modulb)\, \matter^I  \matter^J\
+\ {\rm H.c.} \right)
+\ \cdots,
\eqn\Kexpansion
$$
where $\kappa^2 = {8\pi\over \mpl^2}$ and the $\cdots$
stand for the higher-order terms;
$Z_{\Ib J}$ is the normalization matrix for the observable superfields;
the normalization matrix for the moduli is given by
$\kappa^{-2}\Kmod_{\ib j}\equiv\kappa^{-2}\bar\partial_\ib \partial_j \Kmod$.
In string theory, all terms in \Kexpansion\ suffer from perturbative
corrections,
but in the effective field theory applicable below $\mpl$, only the $Z_{\Ib J}$
matrices renormalize, while the perturbative corrections to the other
parameters
are suppressed.
Corrections to $K$ induced by the hidden sector are similarly suppressed.

Finally, there are moduli-dependent effective gauge couplings
$g_\ind(\modul,\modulb)$ for the different factors in the gauge group
($\Gobs= \prod_\ind G_\ind$).
These couplings renormalize in field theory and are also subject to string-loop
corrections at the string threshold.
However, both effects can be summarized to all orders
in an exact algebraic equation\refmark{\danomaly,\KL}
$$
\!\eqalign{
 g_\ind^{-2}(\modul,\modulb,p)\ ={}\ &
\Re f_\ind(\modul)\ +\ {\cren_\ind\over 8\pi^2}\,\log{\mpl\over p}\
    +\ {\cano_\ind\,\over 16\pi^2}\,\Kmod(\modul,\modulb)\cr
&+\ {T(G)\over 8\pi^2}\,\log g_\ind^{-2}(\modul,\modulb,p)\
    - \sum_r {T_\ind(r)\over 8\pi^2}\,\log\det Z^{(r)}(\modul,\modulb,p)\,
    ,\cr }\!
\eqn\oloopg
$$
where the running of the gauge couplings
and of the $Z$ matrices is explicit ($p$ is the renormalization scale).
\foot{The numerical coefficients in this formula are as follows:
    $\cren_\ind=\sum_r n_r T_\ind(r) - 3 T(G_\ind)$,
    $\cano_\ind =\sum_r n_r T_\ind(r) - T(G_\ind)$,
    $T_{\ind}(r)={\rm Tr}_r(T^2_\ind)$ and $T(G_\ind)=T_\ind({\rm adjoint})$;
    the summation is over representations $r$ of the observable gauge
    group $\Gobs$ and $n_r$ is the number of $\matter^I$ which transform
    like $r$.
    Strictly speaking, eq.~\oloopg\ is exact only for $p\gg\mgrav$.}
The only independent parameters in eq.~\oloopg\ are the holomorphic
functions $f_\ind(\modul)$,
which are closely related to the supersymmetric Wilsonian gauge couplings of
the quantum effective theory (see ref.~\KL\ for details);
at the tree level, $g_\ind^{-2}(\modul,\modulb)=\Re f_\ind(\modul)$.
There are no perturbative corrections to $f_\ind(\modul)$
beyond one loop, although the effective gauge couplings $g_\ind$ are corrected
at all orders because of their dependence on $\Kmod$ and $Z_{\Ib J}$.
Any intermediate-scale threshold corrections to $g_\ind$, \eg\  at $\ms$,
can always be expressed as holomorphic corrections to $f_\ind$.\refmark{\KL}

In order to discuss the implications of
supersymmetry breaking we need to display
the effective potential for the moduli;
neglecting the effects of the electroweak symmetry breaking, we have
$$
\Veff(\modul,\modulb)\ =\ \kappa^{-2}\Kmod_{i\jb}F^i\Fb^\jb\
-\ 3\kappa^2 e^\Kmod |\Wmod|^2 ,
\eqn\pot
$$
where
$$
\Fb^\jb\ =\ \kappa^2 e^{\Kmod/2}\,
\Kmod^{\jb i} \left( \del_i\Wmod + \Wmod\del_i\Kmod\right) \,,\qquad
\Kmod^{\jb i}\,=\,(\Kmod_{i\jb})^{-1}.
\eqn\kderiv
$$
According to our assumptions, at the minimum of eq.~\pot,
$\Veff(\vev\modul,\vev\modulb)=0$,
but (some) $\vev{F^i}\neq0$ and thus supersymmetry
is spontaneously broken.
The measure of this breakdown is the gravitino mass
$$
\mgrav\ =\ \kappa^2 e^{\vev\Kmod/2}\left|\Wmod(\vev\modul)\right|\
=\ \VEV{\coeff13 \Kmod_{i\jb}F^i\Fb^\jb }^{1/2}\
\sim\ {\ms^3\over\mpl^2}
\eqn\mgdef
$$
(the second equation here follows from $\vev{\Veff}=0$);
the magnitude of $\vev{F^i}$ is also  $O(\ms^3/\mpl^2)$.
\foot{We consider the moduli scalars to be dimensionless, which implies
    dimension 1 for the moduli $F$-terms $F^i$.
    All other scalars have dimension 1 and their
     $F$-terms have dimension 2.}

After all the preliminaries, calculating the effective Lagrangian of
the observable sector is quite straightforward.
One starts from the  Lagrangian of the effective theory for
$\matter^I$ and $\modul^i$,
replaces the dynamical moduli fields --- including the auxiliary
$F^i$ --- with their VEVs, and takes the flat limit $\mpl\to\infty$
while keeping $\mgrav$ fixed.\refmark{\BF}
One  finds that the (canonically normalized)
gaugino masses turn out to be\refmark{\MSSM,\IL}
$$
m_\ind\ =\ \coeff{1}{2} F^{i}\del_i\log g^{-2}_\ind \, ;
\eqn\gmass
$$
whereas
the (un-normalized) masses of the observable matter fermions and their
(un-normalized) Yukawa couplings are given by\refmark{\SW,\GM}
$$
\eqalign{
\muh_{IJ}\ &=\ e^{\Kmod/2}\muind_{IJ}\ +\mgrav \Zh_{IJ}\
-\ \Fb^\jb \bar\del_\jb \Zh_{IJ}\, , \cr
\Y_{IJL}\ &=\ e^{\Kmod/2}\,\Ys_{IJL}\,.}
\eqn\fmass
$$
It is convenient to combine both terms into an effective
superpotential
$$
\Weff(\matter)\ =\ \half\muh_{IJ}\matter^I\matter^J\
+\ \coeff13\Y_{IJL}\matter^I\matter^J\matter^L ,
\eqn\weffd
$$
but one should remember that this is a superpotential of the observable sector
and not of the full theory.
In the latter context, \weffd\ would not make sense as a superpotential because
$\muh_{IJ}$ and $\Y_{IJL}$ are non-holomorphic functions of the moduli.
Finally, the potential for the observable scalars (which, by abuse of
notation, we call $\matter^I$) is\refmark{\BF,\SW}
$$
\!\eqalign{
\Veff(\matter,\matterb)\ ={}\ &
\sum_{\ind\in\Gobs}{g_\ind^2\over 4}
    \left( \matterb^\Ib Z_{\Ib J} T_\ind \matter^J\right)^2\
+\ \del_I\Weff Z^{I\Jb} \bar\del_\Jb\overline\Weff \cr
&+\ m^2_{I\Jb}\matter^I \matterb^\Jb\
+\ \left(\coeff13 A_{IJL}\matter^I\matter^J\matter^L\,
    +\,\half\mb_{IJ}\matter^I\matter^J\ +\ \hbox{h.c.}\right) .\cr }\!
\eqn\Vsoft
$$
The first line here gives the scalar potential of an effective theory
with unbroken rigid supersymmetry while
the second line is comprised of the soft
supersymmetry-breaking terms.\refmark{\GG}
The coefficients of these soft terms are as follows:\refmark{\SW}
$$
\eqalign{
m^2_{I\Jb}\ &
=\ \mgrav^2 Z_{I\Jb}\ -\ F^i \Fb^\jb R_{i\jb I\Jb}\,,\cr
A_{IJL}\ &
=\ F^i \Dmod_i \Y_{IJL}\,,\cr
\mb_{IJ}\ &
=\ F^i \Dmod_i \muh_{IJ}\ -\ \mgrav\muh_{IJ}\,,\cr }
\eqn\sresult
$$
where
$$
\eqalign{
R_{i\jb I\Jb}\ &
=\ \del_i \bar\del_\jb Z_{I\Jb}\
    -\ \Gamma_{iI}^N Z_{N\bar L}\overline{\Gamma}^{\bar{L}}_{\jb\Jb}\,,
    \qquad \Gamma_{iI}^N\ =\ Z^{N\Jb} \del_i Z_{\Jb I}\, ,\cr
\Dmod_i \Y_{IJL}\ &
=\ \del_i \Y_{IJL}\ +\ \coeff12 \Kmod_i \Y_{IJL}\
    -\ \Gamma_{i(I}^N \Y_{JL)N}^{}\, ,\cr
\Dmod_i \muh_{IJ}\ &
=\ \del_i \muh_{IJ}\ +\ \coeff12 \Kmod_i \ \muh_{IJ}\
    -\ \Gamma_{i(I}^N\ \muh_{J)N}^{}\,.\cr }
\eqn\defs
$$
(When evaluating $\del_i\muh_{IJ}$ or $\del_i\Y_{IJL}$, one should apply
$\del/\del\modul^i$ to all quantities on the right-hand side of eqs.~\fmass,
 including  $\mgrav$ and $\Fb^\jb$.)
Notice that all quantities appearing in
eqs.~\gmass, \fmass\ and
\sresult\  are  covariant with respect to the supersymmetric
reparametrization of matter and moduli fields
as well as  covariant under \K\ transformations.

According to eq.~\sresult, $m^2_{\Ib J}\sim\mgrav^2$,
$A_{IJL}\sim \mgrav\Y_{IJL}$, and $B_{IJ}\sim\mgrav\muh_{IJ}$;
nevertheless, the soft terms are generally {\it not universal},
\ie\  $A_{IJL}\neq {\rm const}\cdot\mgrav\Y_{IJL}$
and $m^2_{I\Jb}\neq {\rm const}\cdot \mgrav^2 Z_{I\Jb}$,
{\it even at the tree level}.\refmark{\SW}
In the context of the MSSM, this non-universality means
that the absence of flavour-changing neutral currents is not an automatic
feature of supergravity but a non-trivial constraint that has to be satisfied
by a fully realistic theory.\refmark{\IL}

Phenomenological viability of the MSSM imposes  yet another requirement:
The supersymmetric mass term $\muh$ for the two Higgs doublets should be
comparable in magnitude with the non-supersymmetric mass terms.
\foot{Non-minimal supersymmetric extensions of the SM are viable without
    this $\muh$ term, provided the observable sector contains a
    gauge-singlet scalar with Yukawa couplings to both Higgses.\refmark{\EGHRZ}
    (When this singlet gets a VEV, it naturally induces $\muh\sim\mweak$.)
    However, even in non-minimal models one always has to check that
    the Higgses do not become too heavy.}
Equation \fmass\ displays $\muind_{IJ}$ and $\Zh_{IJ}$
as two independent sources of $\muh_{IJ}$,
but one should keep in mind that
$\muind_{IJ}$ depends very strongly
on the couplings of the hidden fields.
Generically, $\muind_{IJ}\sim \ms^n/\mpl^{n-1}\sim \mgrav\cdot(\ms/\mpl)^{n-3}$
(\cf~ref.~\KN),
which means that the cubic and the quartic cross-couplings of the Higgses
to the hidden scalars $(n=1,2)$ can induce unacceptably large
Higgs masses.
\foot{The higher-order cross couplings ($n\ge3$) can also become dangerous if
    $\vev{\hmat^\alpha}\gg\ms$, which happens whenever the hidden sector needs
    non-renormalizable couplings of $\hmat^\alpha$  to prevent
    a $\vev{\hmat^\alpha}\to\infty$ instability.
    Explicit examples of this effect in gaugino condensation
    can be found in  ref.~\KL.}
On the other hand, the contribution of a non-vanishing $\Zh_{IJ}$
to $\muh$ is automatically of order $\mgrav$,  without any fine-tuning.
In our opinion, this  mechanism  --- first pointed out in ref.~\GM\ ---
is
%a better way of
an appealing alternative of
generating the $\muh$-term in the MSSM.

To summarize, we displayed formulae expressing all the couplings of the MSSM
(or, in general, the observable sector) in terms of a few perturbative
parameters of the effective supergravity, namely $\Kmod(\modul,\modulb)$,
$Z_{I\Jb}(\modul,\modulb)$, $\Zh_{IJ}(\modul,\modulb)$,
$\Ys_{IJL}(\modul)$ and $f_a(\modul)$, and even fewer non-perturbative
parameters induced by the hidden sector, namely $\Wmod(\modul)$
and $\muind_{IJ}(\modul)$.
(We have explicitly checked that all higher-order terms
%one could add to
in eqs.~\Kexpansion\ and \Wexpansion\  decouple from the effective Lagrangian
of the observable sector in the limit $\mpl\to\infty$, $\mgrav$ fixed.)
Moreover, the non-perturbative parameters we need to know are holomorphic
functions of the moduli; this gives us a realistic hope for actually computing
those parameters for a large class of hidden sectors.
In addition, not all the observable couplings depend on all the parameters;
for example, $m^2_{I\Jb}$ and the trilinear couplings $A_{IJL}$
are  independent of $\Zh_{IJ}$ and $\muind_{IJ}$.

We would like to stress that eqs.~\gmass, \fmass\ and \sresult\
are valid at all energy scales above $\mgrav$ as long as one uses the
renormalized gauge couplings $g_\ind(p)$ and the renormalized
$Z_{I\Jb}(p)$.
(The other perturbative parameters --- $\Kmod$, $\Zh_{IJ}$ and $\Ys_{IJL}$ ---
do not renormalize below $\mpl$).
We even have an analytic formula~\oloopg\ for the renormalized gauge couplings,
but  all we have for the $Z_{I\Jb}$ matrices are the renormalization group
equations.
Except for a few particularly simple cases, we cannot solve those equations
analytically, whereas the numerical solutions do not give us the moduli
dependence of $Z_{I\Jb}(\modul,\modulb,p)$.
Therefore, in practice one has to use eqs.\
\gmass, \fmass\ and \sresult\ evaluated at the Planck scale,
and then conventional
renormalization group equations  determine the couplings at $\mweak$.
This procedure is quite standard by now and we refer the reader
 to the literature for further details.\refmark{\MSSM}

Nothing we have said so far relied in any way on the stringy nature of the
fundamental theory behind the effective supergravity;
our observations are equally valid for any other unified theory that gives rise
to an effective supergravity below the Planck scale.
However, in the context of string unification, we can make use of the
special properties of the dilaton $S$, which are
common to all $(d=4,N=1)$ vacua of the heterotic string.
All  other moduli (henceforth denoted by $M^i$) we leave completely generic.
At the tree level of the heterotic string,
the couplings of $S$ are universal and are summarized by
$$
\hat{K}^{\rm (tree)}\,=\ -  \log (S + \overline S) \
+\ \Kmod (M,\Mb)\, ,\qquad
f_\ind^{\rm tree}\, =\ k_\ind S\,
\eqn\kdilaton
$$
($k_\ind$ is the level of the relevant Kac--Moody algebra).
Furthermore, $Z_{I\Jb}(M,\Mb)$, $\Zh_{IJ}(M,\Mb)$,
and $\Ys_{IJL}(M)$ are independent of $S$;
their $M$ dependence cannot be further constrained unless one chooses to
focus on a particular class of string vacua.
The non-perturbative
$\Wmod(M,S)$ in eq.~\Whexpansion\ does depend on both $M$ and $S$,
and so do the induced masses $\muind_{IJ}(M,S)$.
However, in string theory  many higher-order couplings often vanish for no
reason that is apparent from the low-energy point of view,
and thus one might expect some string vacua to have
%$\muind_{IJ}=0$ or at least
$\muind_{IJ}\ll\mgrav$.
On the other hand, $\Zh$ is generically non-zero in string theory,
\foot{For the $(2,2)$ string vacua $\Zh_{IJ}$ are related to Yukawa couplings
    of the form  ${\bf 1\cdot 27 \cdot \overline{27}}$
    by the $N=2$ Ward identities analogous to those used in ref.~\DKLa.
    For the $(2,2)$ orbifolds, $\Zh_{IJ}$ vanish in the orbifold limit.}
which presents an alternative mechanism for generating the $\muh$-term.
As far as we know, this possibility has so far been overlooked in the context
of string theory.

At the string loop level, $\Kmod, Z_{I\Jb}$ and $\Zh_{IJ}$
receive an $S$-dependent but generically small threshold correction,
which we neglect in the following discussion.\refmark{\AGNT}
The holomorphic gauge kinetic function $f_a$ is only corrected
at the one-loop level by an $M$-dependent (but $S$-independent)
piece: $f_\ind = k_\ind S\, + f_\ind^{\rm 1-loop} (M)$.
\refmark{\DKLANT,\danomaly}

Generically, the dynamics of the hidden sector can give rise to both
$\vev{F^S}$ and $\vev{F^M}$, but  one type of  $F$-term often dominates over
the
other.
Therefore, we would like to concentrate on the two limiting cases
$\vev{F^S}\gg\vev{F^M}$ and $\vev{F^S}\ll\vev{F^M}$ and  discuss the
phenomenological implications of the two scenarios.
The main feature of the $\vev{F^S}\gg\vev{F^M}$ scenario is the great
simplicity
of the resulting soft terms  before we take  string loops and
 renormalization into account.
Specifically, we find
$$
m_a\ =\ \coeff{\sqrt{3}}{2}\mgrav\,,\qquad
m^2_{I\Jb}\ =\ \mgrav^2 Z_{I\Jb}\,,\qquad
A_{IJL}\ =\ -\sqrt{3}\mgrav\Y_{IJL}\,,
\eqn\sdom
$$
whereas $\muh_{IJ}$ and  $\mb_{IJ}$ are independent parameters.
Thus, in the context of the MSSM the
masses of all super-particles as well as the Higgs VEVs
are determined in terms of the three independent parameters
$\mgrav$, $\muh$ and $\mb$,
and if we further assume that $\muind=0$, then only $\mgrav$ and $\muh$
are independent while $\mb=2\mgrav\muh$.
Numerical study of the electroweak phenomenology produced by these soft terms
shows that for $\muind=0$ the Higgs particle is very light for all allowed
values of the other parameters;
the general case ($\muind\neq 0$) is slightly more involved.
Both cases will be presented in detail in ref.~\BLM.

When the dominant non-perturbative effect in the hidden sector of a
string-based supergravity is the formation of gaugino condensates
  (although other condensates may also be present),
the resulting effective $\Wmod(S,M)$
is more likely to give rise to $\vev{F^M}$
 than to $\vev{F^S}$.\refmark{\dindrsw,\FILQ,\Mbreak}
However, the analysis of this scenario  is much more model-dependent
since the  $M$-dependence of various couplings is quite different
for different string vacua;
nevertheless, even without choosing a particular vacuum
it is possible to make some (interesting) generic statements
about the soft terms.\refmark{\FILQ-\CCMb}
First of all, the usual assumption of the universality of the soft terms in
the MSSM does not automatically hold in this case:
$m^2_{I \Jb}$ is not flavour-blind or even generation-blind;
instead, we have a non-universality parametrized by the field-space
curvature $R_{i\jb I\Jb}$ (see eqs.~\sresult), which generically does not
vanish.
The absence of flavour-changing neutral currents
imposes strong phenomenological constraints on this curvature term
and thus on string model building.\refmark{\IL}
Equations \sresult\ also reveals that the trilinear couplings
$A_{IJL}$ are not strictly proportional to the Yukawa couplings $\Y_{IJL}$,
nor is $\mb_{IJ}$ proportional to $\muh_{IJ}$.

Despite the lack of universality in the $\vev{F^M}$-driven scenario,
we can still make an order-of-magnitude estimate of the supersymmetry-breaking
masses
and couplings.
Barring unexpected cancellations between the two terms in the first
eq.~\sresult,
\foot{One has $m^2_{I\Jb}=0$ whenever  $Z_{I\Jb } = \delta_{I\Jb}e^{\Kmod/3}$,
    or equivalently $R_{i\jb I\Jb} =\coeff13 Z_{I\Jb}\Kmod_{i\jb}$,
which is a feature of  no-scale models.\refmark{\noscale}}
the scalar masses are typically  $O(\mgrav)$. Similarly, the trilinear
couplings $A_{IJL}=$ $O(\mgrav\Y_{IJL})$.
On the other hand, because the gauge couplings depend on the dilaton $S$
more strongly than on the other moduli $M^i$, the gaugino masses come out
rather
light,  $O({\alpha\over4\pi}\mgrav)$ (see eq.~\gmass);
this was first noticed in ref.~\FILQ.
Furthermore, eq.~\oloopg\ allows us to estimate the magnitude of the gaugino
masses after the renormalization, \ie\ just above $\mgrav$.
The result is
$$
m_\ind(p)\ =\ C_\ind\,{\alpha_\ind(p)\over 4\pi}\,\mgrav\ll\mgrav\,,
\eqn\msupp
$$
where the coefficients $C_\ind$ are model-dependent but generally  $O(1)$.
Therefore, in this scenario we expect the gaugino masses to be close
to their experimental lower bounds,
while the squarks and the sleptons should
be around a few TeV.
(For a particular class of models this was indeed  verified in a more careful
numerical analysis in ref.~\CCMb.)
The Higgs mass also tends to be large and the need for a $-M_Z^2$
eigenvalue in the Higgs mass matrix   is another powerful constraint for
string model building.\refmark{\BG}
Finally, $C_a$ in eq.~\msupp\ may differ from gaugino to gaugino,
so in this scenario, the ratio $m_{photino}/m_{gluino}$ may differ
from $\alpha_{QED}/\alpha_{QCD}$.\refmark{\IL}

The two scenarios we  just analysed lead to distinct signals at the weak scale.
We would like to stress  that such signals  do not depend on the
detailed mechanism for supersymmetry breaking nor do they depend on
the chosen string vacuum.
Rather,  they are a mere consequence of  which $F$-term is the dominant seed
of the breaking.
As such they  are not predictions of string theory.

%%%%%%%%%%%%%%%%%%%%%%%%%%%%%%%%%%%%
\smallskip
\noindent \undertext{Acknowledgements}:\quad
The authors would like to thank R.~Barbieri for numerous enlightening
discussions.
We also benefited from discussions  with A.~Casas, M.~Cveti\v c, L.~Ib\'a\~nez,
D.~L\"ust, C.~Mu\~noz, F.~Quevedo, G.~Ross and F.~Zwirner.
\if W\SELECTOR
J.~L. would like to thank the organizers of this workshop
for creating an unusually stimulating environment.
\subpar
The research of V.~K. was supported in part by the NSF (grant \#PHY--9009850)
and by the Robert A.\ Welch Foundation.
%\else
%A preliminary version of this article was presented by J.~L. at the
%Erice Workshop {\it From Superstrings to Supergravity} in December of 1992
%and at the Aspen Winter Conference in January of 1993.
\fi

\refout
\bye